\documentclass[journal=nalefd,manuscript=letter]{achemso}

\usepackage{chemformula} 
\usepackage[T1]{fontenc} 
\usepackage{subcaption}
\author{Weilu Gao}
\email{weilu.gao@utah.edu}
\affiliation[University of Utah]
{Department of Electrical and Computer Engineering, Salt Lake City, Utah}
\author{Cunxi Yu}
\affiliation[University of Utah]
{Department of Electrical and Computer Engineering, Salt Lake City, Utah}
\author{Ruiyang Chen}
\affiliation[University of Utah]
{Department of Electrical and Computer Engineering, Salt Lake City, Utah}

\title[An \textsf{achemso} demo]
  {Artificial Intelligence Accelerators based on Graphene Optoelectronic Devices}


\begin{document}

\begin{tocentry}





\includegraphics{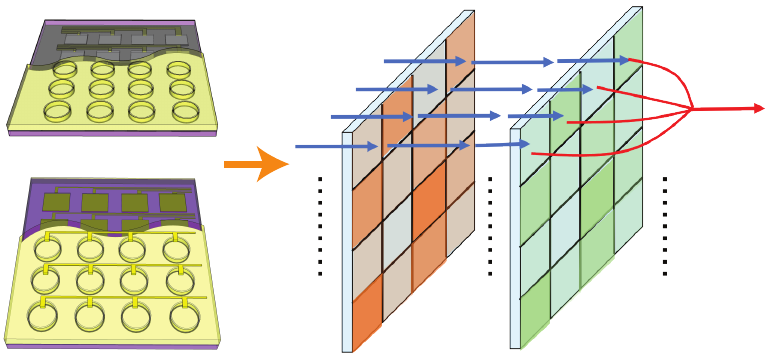}

\end{tocentry}

\begin{abstract}
  Optical and optoelectronic approaches of performing matrix-vector multiplication (MVM) operations have shown the great promise of accelerating machine learning (ML) algorithms with unprecedented performance. The incorporation of nanomaterials into the system can further improve the performance thanks to their extraordinary properties, but the non-uniformity and variation of nanostructures in the macroscopic scale pose severe limitations for large-scale hardware deployment. Here, we report a new optoelectronic architecture consisting of spatial light modulators and photodetector arrays made from graphene to perform MVM. The ultrahigh carrier mobility of graphene, nearly-zero-power-consumption electro-optic control, and extreme parallelism suggest ultrahigh data throughput and ultralow-power consumption. Moreover, we develop a methodology of performing accurate calculations with imperfect components, laying the foundation for scalable systems. Finally, we perform a few representative ML algorithms, including singular value decomposition, support vector machine, and deep neural networks, to show the versatility and generality of our platform. 
\end{abstract}

\section{Introduction}
The past half-decade has seen unprecedented growth in machine learning (ML) algorithms and their applications. For example, deep neural networks (DNNs) represent the state-of-the-art performance in a variety of context, such as large-scale computer vision, natural language processing, and data mining. DNNs have also impacted practical technologies such as web search, autonomous vehicles, and financial analysis~\cite{LeCunEtAl2015N}. However, most of ML algorithms have substantial computational and memory requirements, which greatly limit their training and deployment in resource-constrained environments. To address these challenges, there has been a significant trend in building high-performance specific hardware platforms, such as field-programmable gate arrays~\cite{ZhangEtAl2015,SharmaEtAl2016} and application-specific integrated circuits~\cite{AkopyanEtAl2015ITCDICS,ChenEtAl2016IJSC}. However, with the end of Dennard scaling and Moore's law, the power consumption and density of integrated electronic circuits have hit a bottleneck of processing more complex ML algorithms, especially when state-of-the-art data structures and the number of arithmetic operations are growing from a scale of millions to trillions. For example, the inference of one high-dimension image on a ResNet-50~\cite{HeEtAl2016} requires 3$\times10^{11}$ floating point operations per second (FLOPS), and a single training epoch requires 3$\times10^{13}$ FLOPS to update 25 million parameters. 

Recent efforts on leveraging emerging techniques for efficient ML hardware focus on accelerating the key tensor-level multiply-accumulation (MAC) operations in ML algorithms, which is known as the most computation-intensive operations. For example, analog DNN hardware focuses on accelerating matrix multiplication, such as matrix-vector multiplying module~\cite{SchlottmannEtAl2011IJESTCS}, mixed-mode MAC unit~\cite{WangEtAl2018NE,BankmanEtAl2018IJSC}, and memristor-based MAC~\cite{WangEtAl2017NM,BoybatEtAl2018NC,HuEtAl2018AM}. On the other hand, all-optical and hybrid optoelectronic implementations in early works have offered promising alternative routes to microelectronic implementations~\cite{GoodmanEtAl1978OL,HopfieldEtAl1982PNAS,PsaltisEtAl1985OL,LuEtAl1989AO,DunningEtAl1991OL,ReckEtAl1994PRL,Bar-TanaEtAl1995OL}, because of the advantages of executing MAC operation at the speed of light, high throughput, and very low or even nearly zero power consumption. Recently, an integrated nanophotonic processor based on reconfigurable Mach-Zehnder interferometers at telecommunication wavelength demonstrated the advantages of optical DNN acceleration~\cite{ShenEtAl2017NP,HarrisEtAl2018O}, where MVM operation was decomposed into a series of multiplications following singular value decomposition. Moreover, multiple 3D-printed diffractive optical layers in the terahertz range~\cite{LinEtAl2018S} have shown the capability of performing linear classification, although they are not reconfigurable for new models since weights are physically hardcoded in passive diffractive layers.

In this article, we report a new high-performance optoelectronic architecture of performing general MVM operations by exploiting the extraordinary properties of graphene. Specifically, the architecture consists of a two-dimensional (2D) array of spatial light modulators (SLMs) and a 2D array of photodetectors with electrically controllable photoresponse, which are both constructed out of the combination of large-scale graphene monolayers and optical metamaterials. Since graphene is gapless, these optoelectronic devices can be tailored to operate in ultrabroad frequency range. Considering inevitable non-uniformity of material properties and associated device variation, especially for large-scale polycrystalline graphene, we evaluate the influence of various contributing factors and conceive a methodology of performing accurate calculation even with imperfect devices and systems. Finally, we demonstrate a few representative ML algorithms showing the versatility and generality of the hardware platform. 

\section{Results and discussion}

\begin{figure}
  \includegraphics[width = 0.9\textwidth]{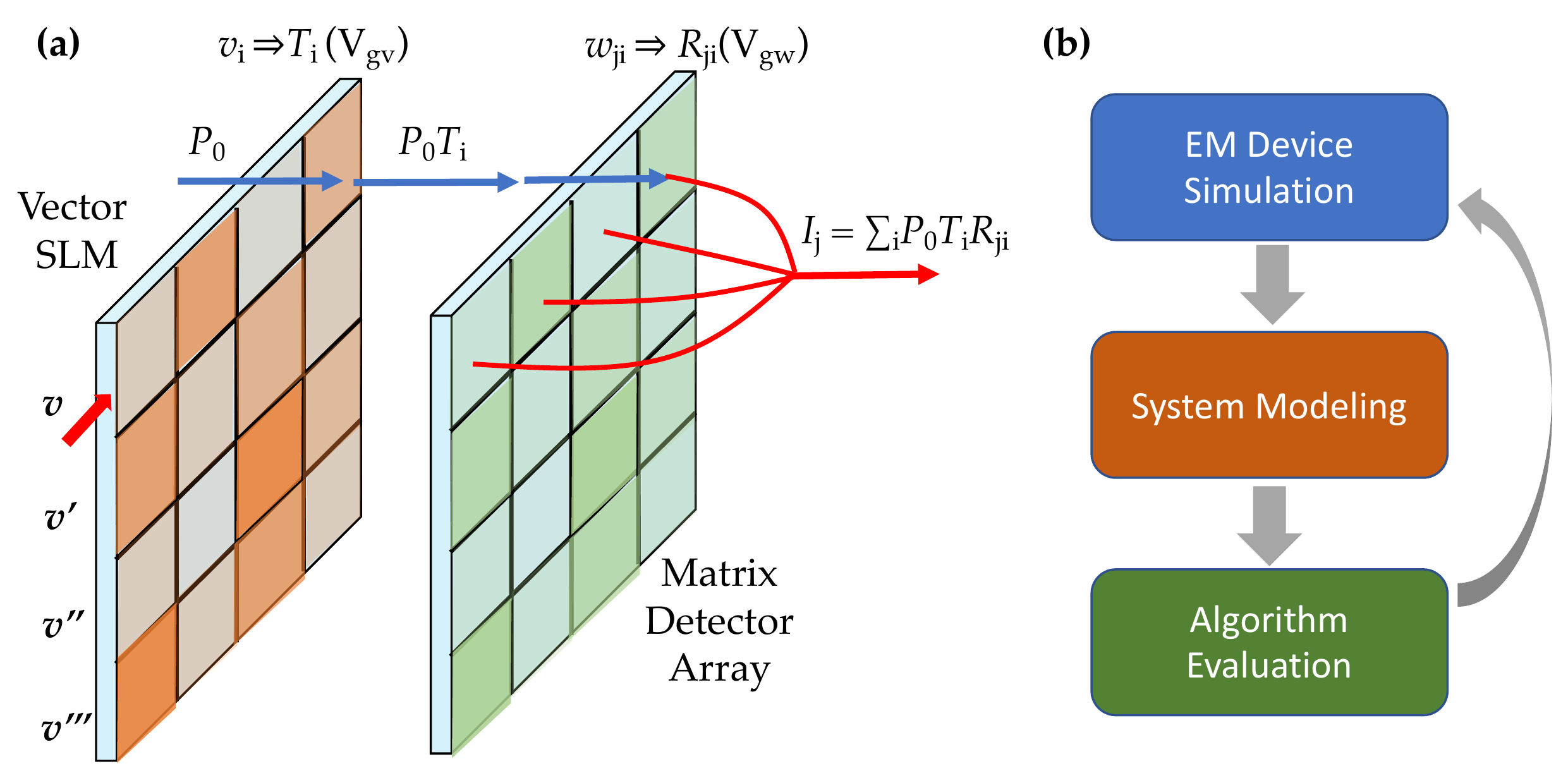}
  \caption{(a)~Schematic of graphene-based optoelectronic matrix-vector multiplier. The vector information is replicated in the 2D array of SLMs and the generated photocurrents from the same row across different columns are added electronically and fed into consecutive circuits for nonlinear operations. (b)~The flowchart for system analysis, including full EM simulations of materials and devices, device abstraction and system modeling, and algorithm benchmarking and evaluation.}
  \label{overview}
\end{figure}

Figure~\ref{overview}a shows an illustration and operation principle of designed architecture to perform a general MVM operation $\vec{o} = {\bf W}\vec{v}$, consisting of a 2D array of SLMs for encoding vector information and a photodetector array with tunable photoresponsivity for encoding matrix element information. The input light is incoherent, such as narrow-band illumination from a halogen lamp with wavelength selection components, so that any coherent inference effect is not involved. An $N$-dimensional vector $\vec{v} = (v_1, v_2, ..., v_N)$ is mapped onto one row of SLMs, and the vector information is also replicated on other rows. This replication has two advantages: (1)~it removes the necessity of involving beam splitting components that restrict chip integration and complicate optical alignment; (2)~it relaxes the requirements of high-quality devices with large-scale uniformity and has a large tolerance of device variation. Each electro-optic unit of SLMs has an electrically-controllable optical transmission function $T_i(\textrm{V}_{\text{gv}})$ encoding the information of $v_i$, and the input power $P_0$ is modulated to $P_0T_i(\textrm{V}_{\text{gv}})$ after the passage. Afterwards, the modulated light is detected by an array of photodetectors, where the photoresponsivity of each element can be electrically controlled. Each element $w_{ji}$ in the matrix {\bf W} is encoded corresponding to the photoresponsivity of a photodetector in the array $R_{ji}(\textrm{V}_{\text{gw}})$. As a result, the obtained photocurrent $I_{ji}$ is $P_0T_i(\textrm{V}_{\text{gv}})R_{ji}(\textrm{V}_{\text{gw}})$, and eventually generated photocurrents are added across columns in the same row that will be converted to voltage for nonlinear activation function implemented using electronic circuits. Mathematically, each element $o_{j}$ in $\vec{o} = (o_1, o_2, ..., o_N)$ corresponds to $\Sigma_{i}I_{ji} = \Sigma_{i}P_0T_i(\textrm{V}_{\text{gv}})R_{ji}(\textrm{V}_{\text{gw}})$. Physically, both optical intensity and readout photocurrent are always-positive values. In order to perform mathematical calculations having both negative and positive real numbers, each element $v_i$ in $\vec{v}$ and $w_{ij}$ in matrix {\bf W} can be represented as a difference of two positive values $v_i = v^{+}_i - v^{-}_i$ and $w_{ji} = w^{+}_{ji} - w^{-}_{ji}$. Thus, the MVM can be done through four multiplications $\vec{o} = {\bf W^{+}}\vec{v^{+}} + {\bf W^{-}}\vec{v^{-}} - {\bf W^{+}}\vec{v^{-}} - {\bf W^{-}}\vec{v^{+}}$. 

In addition, we lay out a design flowchart including electromagnetics (EM) simulation, system abstraction and integration, and performance benchmarking and evaluation; see Fig.\,\ref{overview}b. This flowchart can be generalized to future design and optimization of other architectures. Specifically, the EM simulations connect material optical and optoelectronic properties with device level response; the system modeling incorporates individual device input-output relation to construct a high-level computing architecture with sufficient software interface; ML algorithms utilize such interface to run and evaluate performance, and in turn guide a better material and device design in the bottom level. This ``closed-loop'' style design features an ultimate picture of ``computer-designed computer'' in the future. In this article, we employ such design methodology to demonstrate a graphene-based matrix-vector multiplier operating in the \emph{midinfrared} range, where there have seen many applications, such as thermal imaging~\cite{VollmerEtAl2017} and chem/bio-molecular sensing~\cite{MizaikoffEtAl2013CSR}. 

\begin{figure}
  \includegraphics[width = 0.9\textwidth]{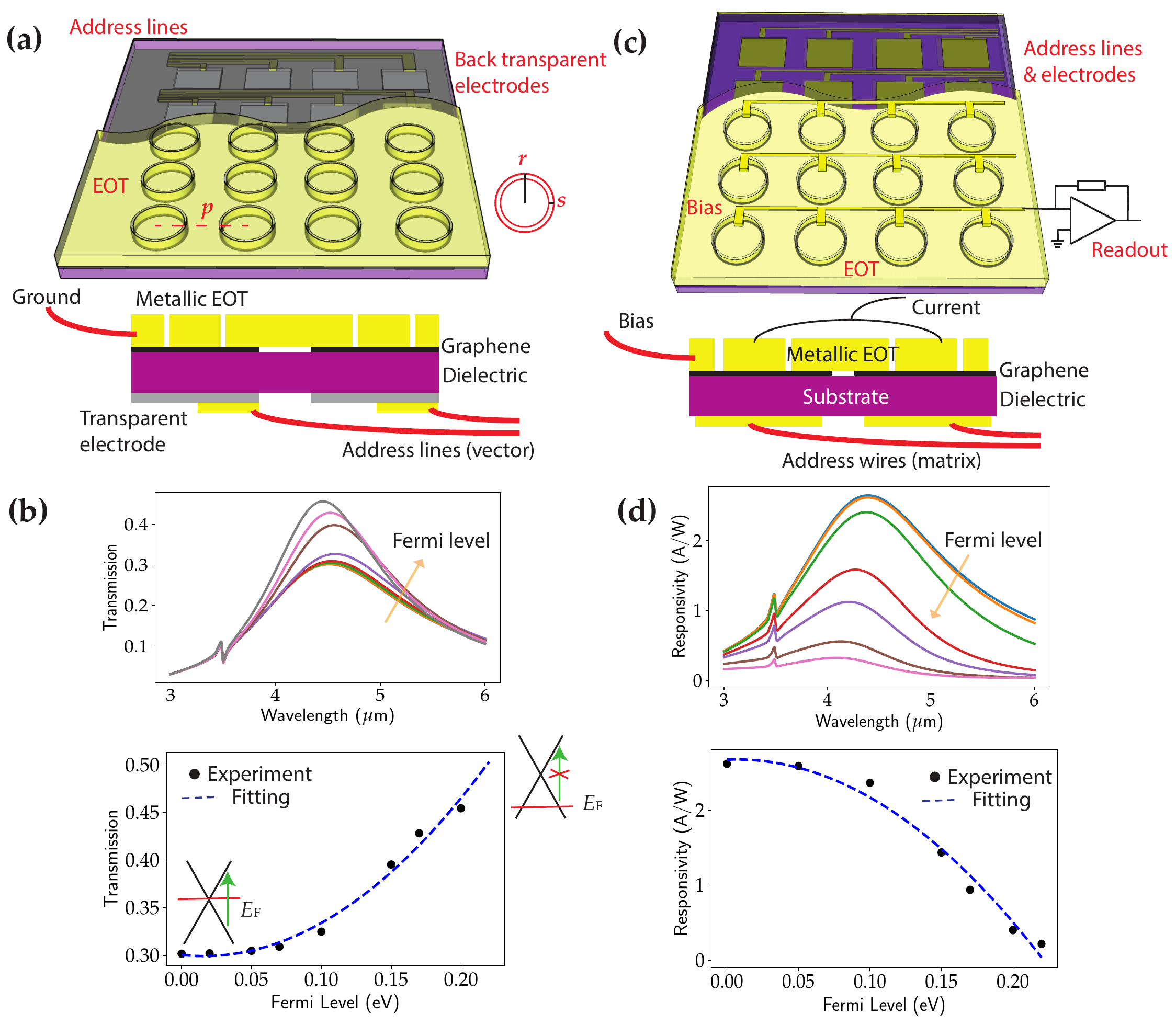}
  \caption{Implementation of graphene-based SLMs and tunable-responsivity photodetectors. (a)~Schematic of active graphene-loaded EOT metamaterials for electro-optic transmission modulation. The EOT metamaterial unit has a 340\,nm outer radius ($r$), a 50\,nm gap ($s$), and the periodicity ($p$) of the array is $1\,\mu$m. Large-scale graphene film is underneath and sits on top of a dielectric layer. Individual electrodes and address wires are used to independently control unit modulator. (b)~Transmission spectra and transmission values at $4.5\,\mu$m of graphene-loaded EOT metamaterials at various $E_{\text{F}}$. Fermi level smaller than half of photon energy allows interband transition while larger Fermi level prohibits the transition. (c)~Schematic of graphene photodetector array, where the responsivity is enhanced by the same EOT metamaterials. A constant bias is applied between outer and inner metal pieces of EOT and the inner metals of structures along the same row are electrically connected for addition operation. The collected current can be converted and amplified to voltage for next-stage nonlinear function processing. (d)~Photoresponsivity spectra and values at $4.5\,\mu$m of graphene-loaded EOT photodetectors at various $E_{\text{F}}$.}
  \label{dev_chara}
\end{figure}

The detailed implementation and characterization of the array of SLMs and photodetectors are summarized in Fig.\,\ref{dev_chara}. Figure \,\ref{dev_chara}a illustrates the design of graphene-based SLMs, which consists of a monolayer graphene and an extraordinary optical transmission (EOT) metamaterial on top~\cite{ShuEtAl2014OE,GaoEtAl2014NL}. The EOT metamaterial unit has a 340\,nm outer radius ($r$), a 50\,nm gap ($s$), and the periodicity ($p$) of the array is $1\,\mu$m. The resulting transmission resonance is positioned around $\approx4.5\,\mu$m. The graphene layer sits on the top of a dielectric thin layer that acts as an insulating layer for electrostatic doping to control the graphene Fermi level ($E_{\text{F}}$) and modify its optical properties. The scattering rate of graphene is assumed to be 2\,meV, corresponding to a carrier mobility $\approx6500\text{cm}^2/(\text{V}\cdot\text{s})$ when $E_{\text{F}} = 0.5$\,eV; see Methods for detailed conversion. Large-scale graphene films of such quality can be readily obtained nowadays using chemical vapor deposition (CVD)~\cite{BanszerusEtAl2015SA}. The EOT array serves dual purposes; one is to enhance the light-matter interaction in graphene so that the modulation efficiency can be significant and the second is to be used as a top electrode for electrostatic control. Underneath each pixel is a transparent electrode, such as nickle ultrathin films~\cite{GhoshEtAl2009OL} and carbon nanotube thin films~\cite{WuEtAl2004S} in the midinfrared range, for addressing each unit modulator. 

Full EM simulation results obtained from commercial software Lumerical FDTD, shown in Fig.\,\ref{dev_chara}b, display the tunable device transmission at various $E_{\text{F}}$ from 0\,eV to 0.22\,eV. When the photon energy (0.27\,eV for 4.5\,$\mu$m) is greater than 2$E_{\text{F}}$, interband transition is allowed with substantial optical absorption and is further enhanced by EOT resonance structure. While when $E_{\text{F}}$ is greater than half the photon energy, due to Pauli blocking, the absorptive transition is forbidden rendering larger transmittance. In a simple parallel-plate capacitor model, $E_{\text{F}}$ is proportional to the square root of gate bias. The accurate relation between gate voltage and Fermi level is strongly dependent on graphene quality, uniformity, and electric gating circuitry. However, such relation is not crucial practically, since the relation between transmission and gate bias is of central interest and can be experimentally determined. The relationship between transmission at 4.5\,$\mu$m wavelength and $E_{\text{F}}$ is fit through a parabolic function and employed in system modeling incorporating device variation; see Methods for more details. Note, in practice, such fitting is not necessary and a look-up table for each device can be used to retrieve applied gate voltage for given output response. 

Similarly, we designed graphene photoconductive detectors as shown in Fig.\,\ref{dev_chara}c, also consisting of EOT metamaterials on top of a monolayer graphene. Given a constant bias between inner and outer metals of EOT structures, the generated photocurrent and photoresponsivity are electrostatically controlled by bottom individual electrode and address wire. All inner metals of EOT structures are connected together to harvest currents from each pixel along the same row to implement addition operation. The collected current can be converted to voltage and amplified for next-stage processing, such as implementing nonlinear activation function. The electrically controllable Fermi level and Pauli blocking switch tune the absorption in graphene. Ideally, if we assume one photon generates a pair of electron and hole that both contribute to measured photocurrent, Figure\,\ref{dev_chara}d demonstrates the photoresponsivity spectra and values at $4.5\,\mu$m under various Fermi levels, and the latter is also fit using a parabolic function. In opposite to transmission, the responsivity decreases with increasing Fermi levels because of blocked interband absorption. 

In contrast to electronic implementation, this graphene optoelectronic architecture has ultrahigh parallelism, where all elements in both vector and matrix are computed simultaneously. In addition, the ultrahigh carrier mobility of graphene promises fast switching speed for the implementation of both SLMs and photodetectors, which can be readily above 1\,GHz to even tens of GHz~\cite{PhareEtAl2015NP}. These two factors suggest the ultrahigh data throughput of the system. Furthermore, the electrostatic control and tuning of both SLMs and photodetectors are especially power efficient -- with nearly zero power consumption -- in the static state and inference mode. This architecture also has potential of being integrated into a single chip, thanks to the large-scale CVD growth of graphene and its compatibility with modern micro/nanofabrication processes. 

\begin{figure}
  \includegraphics[width = 0.9\textwidth]{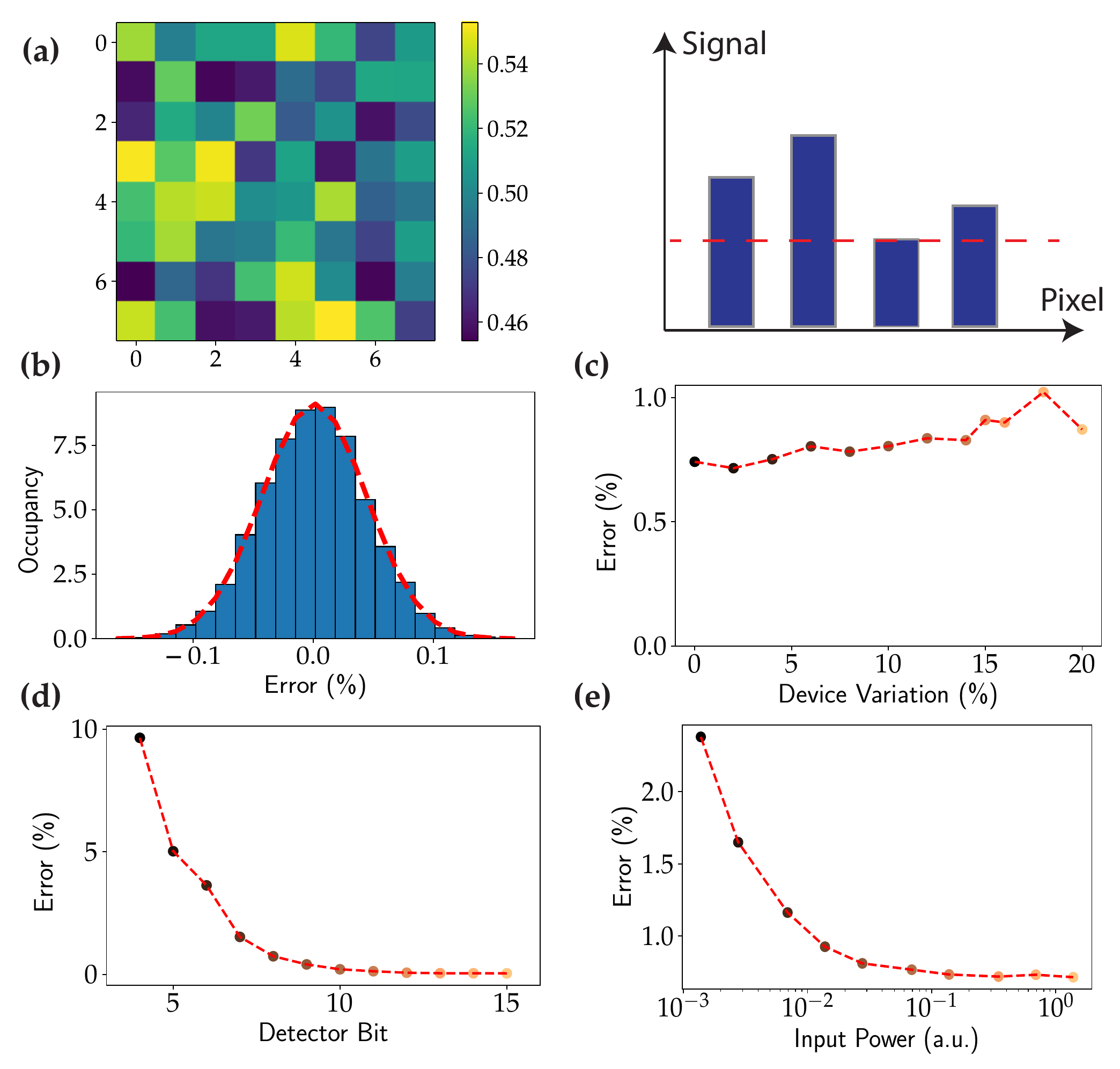}
  \caption{System analysis with imperfect components. (a)~The transmittance spatial distribution for an $8\times8$ SLM array with $20\%$ variation and the methodology of correcting such variation in a row by choosing the minimum tuning range of SLM and detector pair. (b)~A representative error distribution by comparing the calculation results from graphene multiplier with linear algebra calculation. The distribution is fit using a Gaussian function and a standard deviation is used as the figure of merit. (c) -- (e)~The figure of merit standard deviation error as a function of device variation, detector bit precision, and input power.}
  \label{accuracy}
\end{figure}

One important issue of emerging architectures bearing photonic and generally analog computing is scalability, which is especially notorious involving unconventional materials, such as graphene nanomaterials. In current example, there are inevitable device variation and non-uniformity when the array scale is large, which can be due to polycrystalline nature of graphene and micro/nanofabrication variation. Performing accurate calculation with imperfect components is thus crucial for practical deployment, and the procedure of correcting such imperfection is necessary. Figure\,\ref{accuracy}a shows an $8\times 8$ array of SLMs with 20\% transmission variations at the same graphene Fermi level or gate voltage and similar variation also occurs in the responsivity of photodetector array; see Methods for details about how the strength of variation is added onto device parameters. 

In our correction procedure, for each row, we sweep applied gate voltages on each unit of SLMs and corresponding photodetectors pair by pair, and for each pair, we sweep gate voltage of SLM unit and that of photodetector unit separately. From the readout, we obtain tuning curves for each unit of SLMs and photodetectors. Due to the non-uniformity of devices, the tuning range from each pair can vary, as illustrated in Fig.\,\ref{accuracy}a. The developed strategy is to define the \emph{minimum} tuning range on that row as the physical quantity unit so that any other reading from the row readout can be converted to algebraic values by dividing this unit. Also, defining the unit as the minimum tuning range guarantees that each pair can achieve this range. This methodology highlights the advantages of replicating vector encoding on the rows of SLMs, through which the correction for each row is independent from others. In contrast, the calibration in the structures involving beam splitting elements is cross-linked between rows and is significantly more complicated. The detailed mathematical analysis and proof of correction procedure to generate the accurate output results are provided in \emph{Supporting Information} Section 1. 

The accuracy of graphene multiplier is evaluated by comparing the calculation results with those obtained from standard linear algebra multiplication function. Figure\,\ref{accuracy}b shows a representative calculation error distribution of 10000 multiplication calculations of a random $8\times8$ matrix and a random $8\times1$ vector with all elements within $-1$ to 1. The histogram is fit using a normal distribution and the standard deviation is the figure of merit for evaluation. Figure \,\ref{accuracy}c displays the standard deviation error for various degrees of device variation from 0 to $20\%$. This variation applies to both SLMs and photodetectors. The error is nearly constant by using the correction procedure described above, proving the effectiveness of this procedure. Note here, the residual error for perfect devices originates from the finite precision of applied gate voltage, which is assumed to be 8 bit. In addition to the limit of finite precision in applied gate voltages, the readout from detectors can also have finite precision. For example, commercially available digital CCD cameras in the visible range generally have 10 bit precision. We also investigated the influence of detector bit precision on accuracy performance, and as shown in Fig.\,\ref{accuracy}d, the error drastically increases with small bit precision (e.g. 5 bit). Finally, we investigated the influence of noise in the system, which is modeled as a Gaussian noise added onto the readout end. The noise effect is reflected onto the error dependence on input power. Note here, the detector responsivity has been ideally modeled and in practice the responsivity can be quite different. Thus, the unit of input power on the x-axis is arbitrary unit. As expected, as the input power and thus signal-to-noise ratio decreases, the error increases. More error histograms for these contributing variations and noise are provided in \emph{Supporting Information} Section 2. 

\begin{figure}
  \includegraphics[width = 1.0\textwidth]{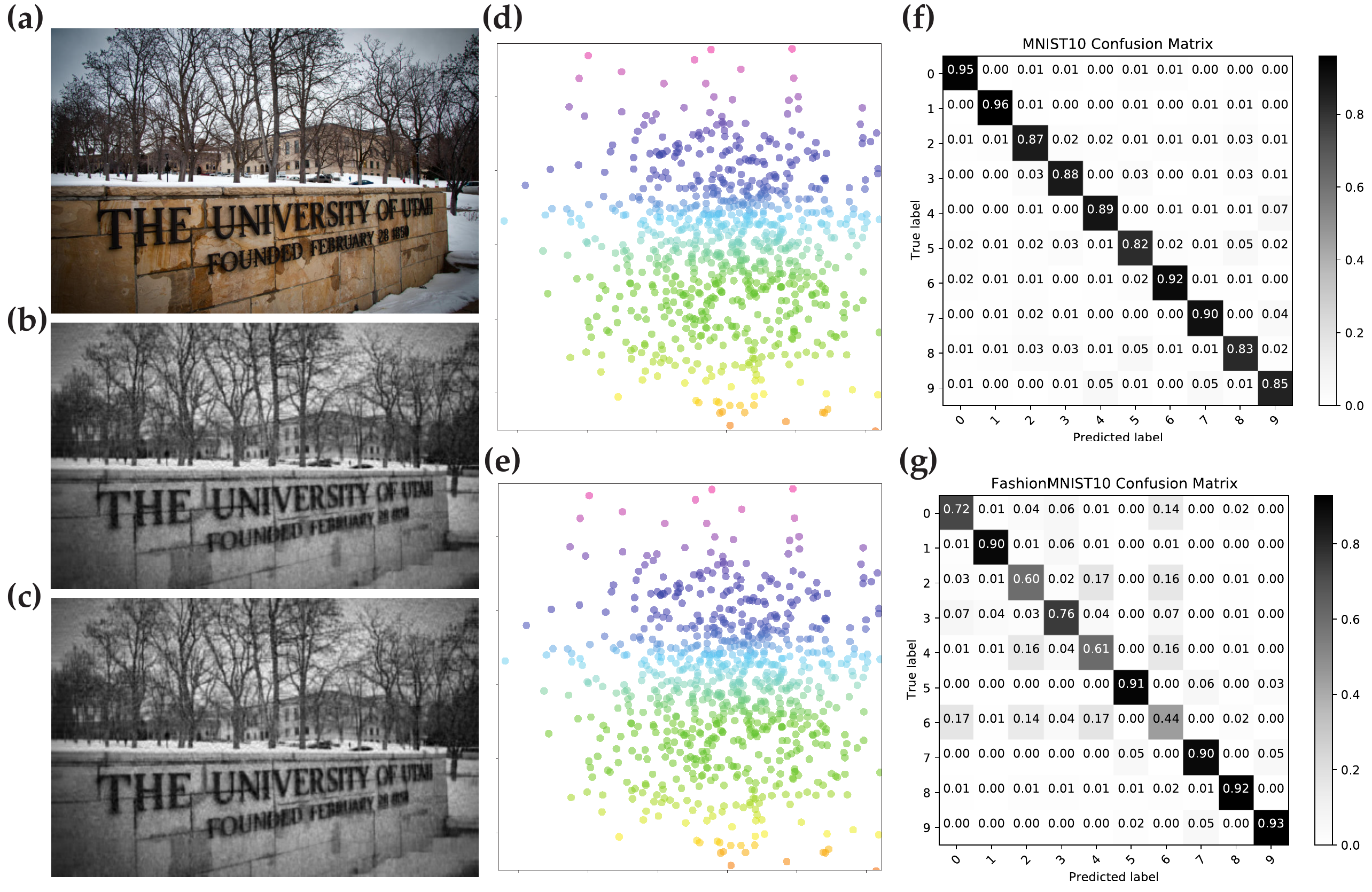}
  \caption{ML algorithms demonstration. (a)--(c)~Demonstration of singular value decomposition, including (a)~original image, (b)~reconstructed image with $K=50$ using GPPs, and (c)~reconstructed image with $K=50$ using graphene architecture. (d),(e)~Demonstration of support vector machine, including (d)~the clustering of \texttt{Blobs} dataset with loss 0.0626 obtained using GPPs and (e)~the clustering of the same data with loss 0.0627 obtained using graphene architecture. (f),(g)~Demonstration of neural networks for the classification of (f)~\texttt{MNIST10} dataset with 88.7\% accuracy and (g)~\texttt{Fashion-MNIST10} dataset with 76.8\% accuracy.}
  \label{algorithm}
\end{figure}

Finally, we utilize our graphene multiplier for running multiple ML algorithms. We emulated and corrected an $8\times8$ multiplier, and established a general matrix-matrix multiplication (GEMM) by segmenting the matrix into multiple blocks to fit the dimension of our multiplier emulator; see Methods for more details. We compare the quality of results of selected ML algorithms obtained with our GEMM multiplier with the results from a general-purpose processor (GPP), which is an Intel Xeon Gold 6230 processor in this work. First, we evaluated the graphene GEMM for image reconstruction, in which the image was compressed using \textit{singular value decomposition} (SVD). The original image is shown in Fig.\,\ref{algorithm}a, and has been compressed using SVD such that $\text{image} = U \cdot \Sigma \cdot V^{T}$, where the dimensionalities of $\text{image}, U, \Sigma, V^{T}$ are $\mathcal{R}^{m \times n \times 3}$, $\mathcal{R}^{m \times p}$, $\mathcal{R}^{n \times p}$, and $\mathcal{R}^{p \times p}$, respectively. Specifically, our experiments were conducted on image $\in \mathcal{R}^{768 \times 512 \times 3}$ ($m=768$, $n=512$). While the top singular vectors capture most of the variation, instead of using all the singular vectors and multiplying them as shown in SVD decomposition, we reconstructed the image using top-$K$ singular vectors. The reconstructed image ($K=50$) with GPPs (Figure\,\ref{algorithm}b) has the same quality as that of image reconstructed using graphene multiplier (Figure\,\ref{algorithm}c). The second ML algorithm we evaluated with graphene GEMM is unsupervised learning using \textit{support vector machine} (SVM) algorithm on \texttt{Blobs} dataset. As shown in Figs.\,\ref{algorithm}d and e, the clustering results generated with our GEMM multiplier match the results obtained on GPPs, where the loss differs $< 0.2\%$.

Figures\,\ref{algorithm}f and g display another demonstration on ML algorithms conducted on \textit{multi-layer perceptron} (MLP) neural networks. Specifically, we built and trained a two-layer MLP network without using nonlinear activation function for two multiclass image classification datasets, \texttt{MNIST10} and \texttt{Fashion-MNIST10}. Details about training settings can be found in Methods. Figures\,\ref{algorithm}f and g present the prediction confusion matrices for these two datasets, where the graphene multiplier achieved 88.7\% accuracy for MNIST10 and 76.8\% accuracy for Fashion-MINST10. In comparison, the GPP achieved slightly better prediction performance using the same MLP architecture, saying 92.3\% and 78.7\% accuracy for MNIST10 and Fashion-MNIST10, respectively. While we demonstrate that graphene GEMM multiplier can achieve similar results as GPPs for the first two ML algorithms, there are noticeable accuracy degradations for image classification tasks using MLPs. We find out that the accuracy degradations are mainly caused by initialization and training algorithms, such that the learned parameters in MLPs are very small with the mean close to zero. Due to such distribution of MLP parameters, the inevitable errors from graphene multiplier associated with noise and finite precision become noticeable than other applications. However, we believe that the impacts of errors introduced by graphene multiplier will be much smaller while applying to more robust and larger neural network architectures.

In summary, we report a new high-performance optoelectronic architecture of performing general MVM and GEMM operations by exploiting the extraordinary properties of graphene. Specifically, this architecture consists of a 2D array of SLMs and photodetectors with electrically controllable transmission and photoresponse, which are both constructed out of the combination of large-scale graphene monolayers and optical EOT metamaterials. This system possesses ultrahigh data throughput and ultralow power consumption, because of extreme parallelism of the architecture, ultrahigh carrier mobility of graphene, and electrostatic control. From the perspective of practically deploying large-scale system, we design a methodology of performing accurate calculation with imperfect devices and systems and evaluate the influence of imperfection, considering inevitable non-uniformity of material properties and associated device variation. Finally, we demonstrate a few ML algorithms showing the versatility and generality of the hardware.

\newpage
\section{Methods}

\noindent{\bf Graphene model in Lumerical FDTD}

\noindent Graphene monolayer is modeled as a 2D rectangle conducting sheet in Lumerical material library, including both interband and intraband contributions. Fermi level and scattering rate are two parameters used to calculate dielectric constants used for Lumerical. Scattering rate used in Lumerical can be converted to mobility as follows. The damping constant $\gamma = \frac{q\hbar v^2_\text{F}}{\mu E_\text{F}}$ is twice of scattering rate setting in Lumerical library, where $q = 1.6\times10^{-19}$\,C is the value of electron charge, $\hbar = 1.05\times10^{-34}\text{J}\cdot\text{s}$ is the reduced Planck constant, $v_\text{F} = 1\times 10^6$\,m/s is the Fermi velocity, $\mu$ is the carrier mobility, and $E_\text{F}$ is the Fermi level. Thus, in this study, 2\,meV damping rate, corresponding to 1\,meV scattering rate in Lumerical setting, is used and is equivalent to the carrier mobility $\approx6500\text{cm}^2/(\text{V}\cdot\text{s})$ at $E_\text{F} = 0.5$\,eV.

\medskip
\noindent {\bf Device response fitting and variation modeling}

\noindent The simulated transmission of SLMs and absorption of photodetectors using EM simulators as a function of various Fermi levels are fit using 2nd-order polynomials. Concretely, $T(V_g) = a_t^{(2)}V^2_g + a_t^{(1)}V_g + a_t^{(0)}$ and $R(V_g) = b_r^{(2)}V^2_g + b_r^{(1)}V_g + b_r^{(0)}$. The device variation is modeled as that fitting parameters $a_t^{(i)}$ and $b_r^{(i)}$ vary across different devices in the array of SLMs and photodetectors. Specifically, take SLMs for example, the parameter vector $\vec{a} = (a_t^{(2)}, a_t^{(1)}, a_t^{(0)})$ varies as $\vec{\tilde{a}} = -pX\vec{a} + (1.0 + p/2)\vec{a}$, where $X$ is a random number between 0 and 1 with uniform distribution generated for each unit and $p$ denotes the strength of variation. $\vec{\tilde{a}}$ is in the range between $(1.0 - p/2)\vec{a}$ and $(1.0 + p/2)\vec{a}$. 20\% variation means $ p = 0.2$ and for each unit of the array $X$ is randomly generated.

\noindent {\bf Implementation of general matrix multiply (GEMM)}

\noindent GEMM is a common algorithm in linear algebra, machine learning, statistics, and many other domains. Mostly, this includes using blocking, inner products, outer products, and systolic array techniques, which breaks the computations of GEMM to better utilize vector-multiplication or MVM. Specifically, for this work, we develop optoelectronic GEMM by utilizing the proposed optoelectronic MVM, where we decompose the targeted matrices into block-matrices (also known as \textit{block partitioning}). GEMM is then implemented recursively using \textit{divide-and-conquer} algorithm, which is used to execute the ML algorithms discussed in Fig.\,\ref{algorithm}.

\noindent {\bf Training of ML algorithms}

\noindent \textit{Autograd} is a reverse automatic differentiation system, which records a graph representation of all the operations that encode the input-output mappings of ML models. As a result, it returns as a directed acyclic graph whose leaves are the input tensors and roots are the output tensors. By tracing this graph from roots to leaves, gradients can be automatically computed based on the chain-rule for gradient-based \textit{backpropogation} algorithms. While the evaluated ML algorithms are all implemented with GEMM operators, we can simply construct the autograd graphs using \texttt{PyTorch-autograd} mechanism and deploy gradient descent algorithm \texttt{Adam} to train the ML models according to a given loss function. Specifically, we use \textit{mean-square-error} loss to train SVM-based clustering application, and \textit{negative-log-likelihood-loss} to train \texttt{MNIST10} and \texttt{Fashin-MNIST10} classification tasks. Our \texttt{Adam} backpropagation algorithm settings include learning rate $lr=0.1$, $\beta_1=0.9$, $\beta_2=0.999$, $\epsilon=10^{-8}$, and without $L2$ penalty. To evaluate the final prediction performance of those ML algorithms with the proposed optoelectronic GEMM architecture, we replace the PyTorch matrix-multiply functions with our GEMM algorithm.

\begin{acknowledgement}

  W.\,G. thanks the support from the University of Utah start-up fund. Y.\,C. thanks the support from grants NSF-2019336 and NSF-2008144.



\end{acknowledgement}

\begin{suppinfo}

\noindent The Supporting Information is available free of charge.

Information on the mathematical analysis of correction procedures and a few representative calculation error histograms (PDF).

\end{suppinfo}

\bibliography{weilu.bib}

\end{document}